\documentclass[11pt]{article}
\usepackage{graphicx}
\usepackage{wrapfig}

\def\Caption#1{\caption{\small #1}}

\def\Title#1{\begin{center} {\Large #1 } \end{center}}
\def\Author#1{\begin{center}{ \sc #1} \end{center}}
\def\Address#1{\begin{center}{\small \it #1} \end{center}}

\newenvironment{Abstract}{\begin{quotation}  }{\end{quotation}}
\newenvironment{Presented}{\begin{quotation} \begin{center} 
             PRESENTED AT\end{center}%\bigskip 
      \begin{center}\begin{large}}{\end{large}\end{center} \end{quotation}}
\def\Acknowledgements{\bigskip  \bigskip \begin{center} \begin{large}
             \bf ACKNOWLEDGEMENTS \end{large}\end{center}}
\def\beq{\begin{equation}}
\def\eeq#1{\label{#1}\end{equation}}
\def\eeqn{\end{equation}}

\def\beqa{\begin{eqnarray}}
\def\eeqa#1{\label{#1}\end{eqnarray}}
\def\eeqan{\end{eqnarray}}

\let\bar=\overbar

\def\Dslash{\not{\hbox{\kern-4pt $D$}}}
\def\dslash{\not{\hbox{\kern-2pt $\del$}}}

\def\msb{{\bar{\ssstyle M \kern -1pt S}}}

\def\nz{ --- }
\def\deg{$^\circ$}
\def\fref#1{Fig.\,\ref{#1}}

\begin{document}
\begin{titlepage}
%\pubblock

\vfill
\Title{Recent results from the KASCADE-Grande cosmic-ray experiment --- Test of
hadronic interaction models with air-shower data}
\vfill
\Author{J\"org R. H\"orandel$^j$,
W.D. Apel$^a$, J.C. Arteaga-Velázquez$^b$, K. Bekk$^a$, M.~Bertaina$^c$, J.
Bl\"umer$^{a,d}$, H.  Bozdog$^a$, I.M. Brancus$^e$, E. Cantoni$^{c,f,n}$,
A.~Chiavassa$^c$, F. Cossavella$^{d,o}$ K.  Daumiller$^a$, V. de Souza$^g$, F.
Di Pierro$^c$, P.~Doll$^a$, R. Engel$^a$, D. Fuhrmann$^{h,p}$, A.
Gherghel-Lascu$^e$, H.J. Gils$^a$, R.~Glasstetter$^h$, C. Grupen$^i$, A.
Haungs$^a$ D.  Heck$^a$, D. Huber$^d$, T. Huege$^a$, K.-H.~Kampert$^h$, D.
Kang$^d$, H.O.  Klages$^a$, K. Link$^d$, P. Luczak$^k$, H.J. Mathes$^a$, H.J.
Mayer$^a$, J. Milke$^a$, B.  Mitrica$^e$, C. Morello$^f$, J.
Oehlschl\"ager$^a$, S.~Ostapchenko$^l$, N. Palmieri$^d$, M.  Petcu$^e$, T.
Pierog$^a$, H. Rebel$^a$, M. Roth$^a$, H.~Schieler$^a$, S. Schoo$^a$, F.G.
Schr\"oder$^a$, O. Sima$^m$, G. Toma$^e$, G.C.~Trinchero$^f$, H. Ulrich$^a$, A.
Weindl$^a$, J.  Wochele$^a$, J. Zabierowski$^k$}
\Address{
$^a$Institut f\"ur Kernphysik, KIT – Karlsruhe Institute of Technology, Germany\nz
$^b$Universidad Michoacana de San Nicolas de Hidalgo, Inst. Física y Matematicas, Morelia, Mexico\nz
$^c$Dipartimento di Fisica, Università degli Studi di Torino, Italy\nz
$^d$Institut f\"ur Experimentelle Kernphysik, KIT – Karlsruhe Institute of Technology, Germany\nz
$^e$Horia Hulubei National Institute of Physics and Nuclear Engineering, Bucharest, Romania\nz
$^f$Osservatorio Astrofisico di Torino, INAF Torino, Italy\nz
$^g$Universidade São Paulo, Instituto de Fîsica de São Carlos, Brazil\nz
$^h$Fachbereich Physik, Universität Wuppertal, Germany\nz
$^i$Department of Physics, Siegen University, Germany\nz
$^j$Department of Astrophysics/IMAPP, Radboud University Nijmegen, The Netherlands\nz
$^k$National Centre for Nuclear Research, Department of Astrophysics, Lodz, Poland\nz
$^l$Frankfurt Institute for Advanced Studies (FIAS), Frankfurt am Main, Germany\nz
$^m$Department of Physics, University of Bucharest, Bucharest, Romania\nz
$^n$Present address: Istituto Nazionale di Ricerca Metrologia, INRIM, Torino, Italy\nz
$^o$Present address: DLR Oberpfaffenhofen, Germany\nz
$^p$Present address: University of Duisburg-Essen, Duisburg, Germany 
}
\vfill
\begin{Abstract}
Cosmic rays provide an unique approach to study hadronic interactions at 
high energies in the kinematic forward direction. The KASCADE air shower
experiment was the first to conduct quantitative tests of hadronic interactions
with air shower data. A brief overview is given on results from KASCADE and its
extension KASCADE-Grande with respect to investigations of hadronic
interactions and the properties of cosmic rays.
\end{Abstract}
\vfill
\begin{Presented}
 EDS Blois 2017, Prague,  Czech Republic, June 26-30, 2017
\end{Presented}
\vfill
\end{titlepage}
\def\thefootnote{\fnsymbol{footnote}}
\setcounter{footnote}{0}

\section{Introduction}
\begin{wrapfigure}{r}{0.5\textwidth}
 \includegraphics[width=0.5\textwidth]{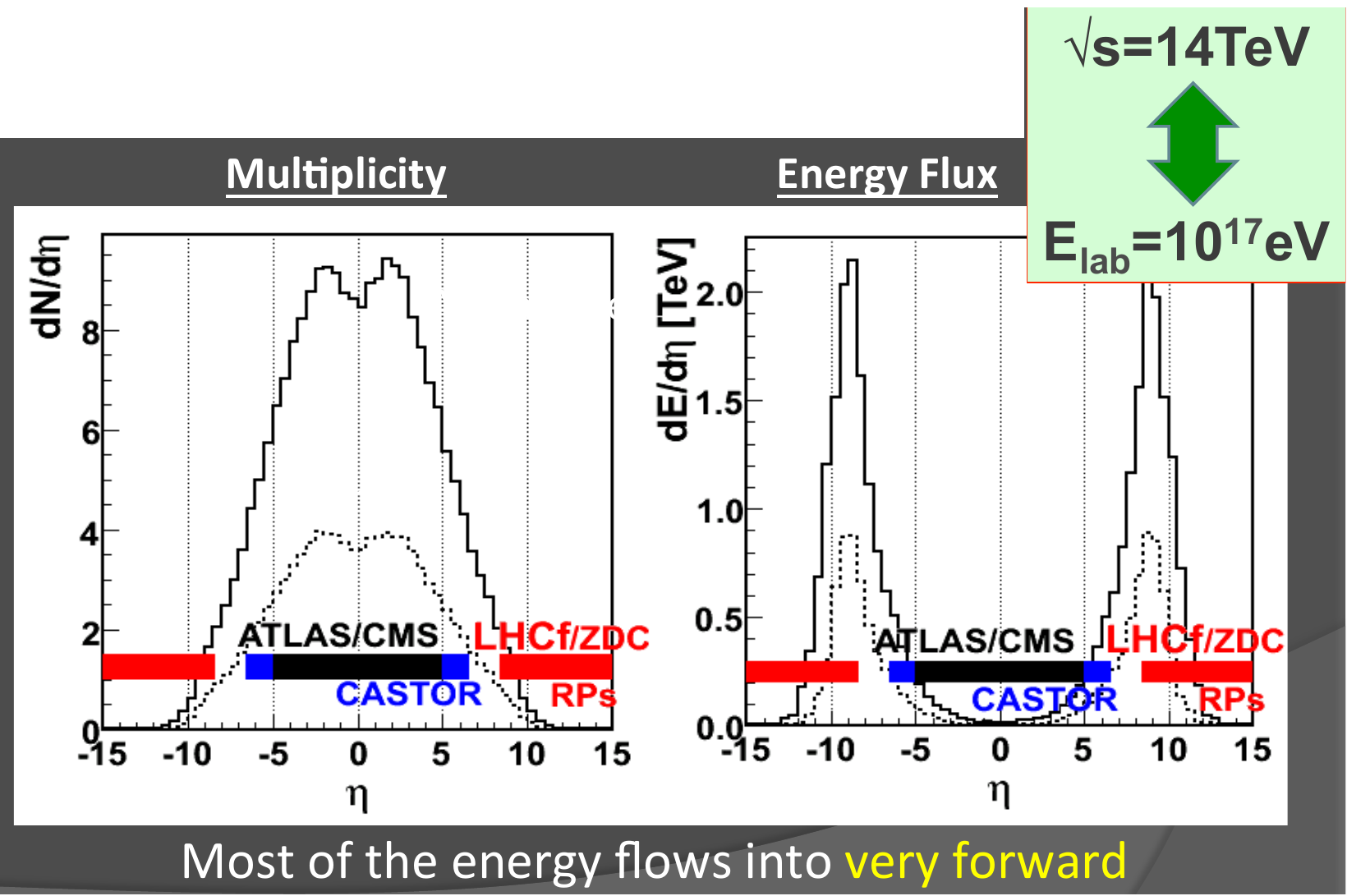}
 \Caption{Multiplicity (left) and energy flux (right) in hadronic interactions
 as a function of pseudorapidity. In addition, the coverage of detectors at the
 LHC is shown. RP refers to Roman Pots, ZDC means zero-degree
 calorimeter.}\label{eta}
\end{wrapfigure}
Cosmic rays are ionized atomic nuclei impinging on the atmosphere from outer
space. Particles with energies exceeding $10^{20}$~eV have been measured, being
the highest-energy particles in
nature\cite{naganowatson,Haungs:2003jv,behreview}.  When they impinge on the
atmosphere they initiate cascades of secondary particles, the extensive air
showers (EASs). The atmosphere with a thickness of $\approx30$ radiation
lengths or $\approx11$ hadronic interaction lengths acts as a calorimeter.
Cosmic rays are an unique probe to study hadronic interactions at energies well
beyond the regime of human made accelerators. The center of mass energy of the
LHC $\sqrt{s}=14$~TeV corresponds to a laboratory energy of $10^{17}$~eV,
relevant for cosmic-ray studies. The multiplicity of secondary particles in
(central) collissions peaks around pseudorapidity% 
  \footnote{The pseudorapidity is defined as
  $\eta=-\ln\left[\tan(\theta/2)\right]$, where $\theta$ is the angle between
  the primary particle (or the beam axis) and the secondary particle.}
values of $\eta=0$. Thus, the main experiments e.g.\ at the LHC such as ATLAS
and CMS cover these central regions up to $\eta\approx\pm 7$, see \fref{eta}.
However, the energy flux exhibits maxima at pseudorapidity values around
$\eta\approx\pm10$.  At the LHC this region is
covered by specialized forward detectors such as LHCf or TOTEM.  The region in
the extreme forward direction $\eta\approx\pm 10$ is of great importance for
the development of extensive air showers.  Thus, cosmic rays and extensive air
showers are used to study hadronic interactions at highest energies in the
kinematic forward direction.

The KASCADE air shower experiment was one of the first experiments to conduct
quantitative studies of hadronic interactions with air shower data.  In the
following, a brief introduction to the KASCADE and KASCADE-Grande experiments
is given. Followed by a review of results on hadronic interactions and the
properties of cosmic rays.

\section{The KASCADE-Grande experiment}
The KASCADE-Grande \cite{Apel:2010zz} detector array was situated at the site
of the Karlsruhe Institute of Technology – KIT, Campus North, Germany (49\deg
N, 8\deg E) at 110 m a.s.l. It had a roughly quadratical shape ($\approx
700\times700$~m$^2$) and it comprised a multi-detector system. Several types of
detectors enabled the measurement of different air-shower observables.
Historically, the KASCADE-Grande detector was an extension of a smaller array,
KASCADE \cite{kascadenim}, which has been operated from 1996 on. KASCADE was a
complex detector system aimed to clarify the origin of the knee in the energy
spectrum of cosmic rays at a few $10^{15}$~eV.  It was designed to measure
cosmic rays with energies between $10^{14}$ and $10^{16}$~eV, by simultaneously
recording the electromagnetic, muonic, and hadronic shower components.
A $200\times200$~m$^2$ scintillator array recorded the electromagnetic and
muonic ($E_\mu>230$~MeV) shower components.  
\begin{wrapfigure}{r}{0.5\textwidth}
 \includegraphics[width=0.5\textwidth]{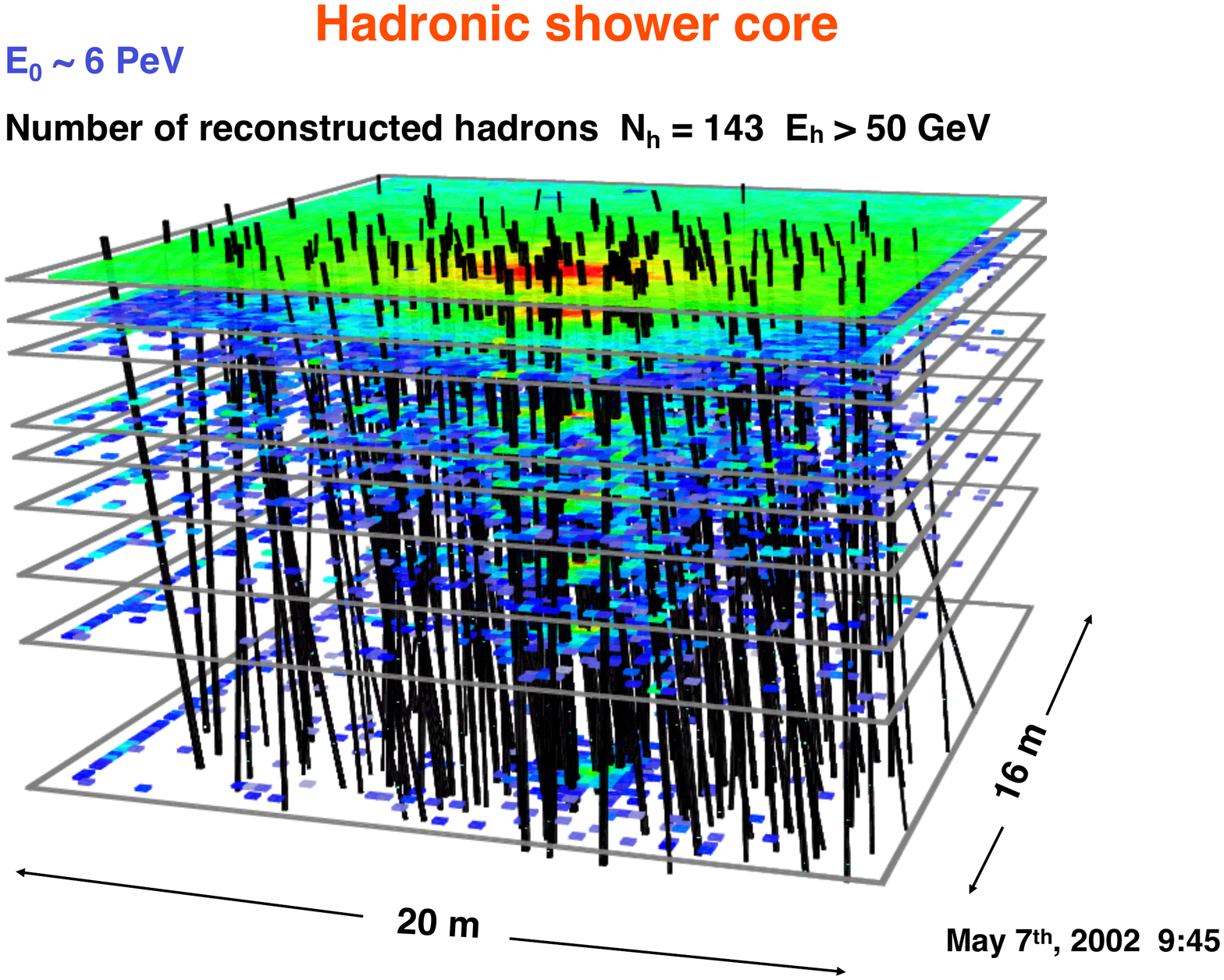}
 \Caption{Hadrons in the core of an air shower with an energy of
 $\approx6\cdot10^{15}$~eV measured with the KASCADE hadron calorimeter. 143
 individual hadrons have been reconstructed with energies exceeding 50 GeV.}
 \label{hadcore}
\end{wrapfigure}
Of particular interest for the study of hadronic interactions was a 320~m$^2$
hadronic sampling calorimeter ($E_h>20$~GeV) with a total depth of 11 hadronic
interaction lengths, interspaced with nine layers of liquid ionization chambers
\cite{kalonim}, see \fref{hadcore}.  To extend the energy range up to
$10^{18}$~eV was the motivation for the extension KASCADE-Grande, thereby
focusing on the expected transition from Galactic to extragalactic cosmic rays
in the energy range $10^{17}-10^{18}$~eV.  The Grande array comprised 37
detector stations with 10 m$^2$ plastic scintillator each (formerly part of the
EAS TOP array \cite{eastop}), which were arranged on a roughly hexagonal grid
with a spacing of about 140~m.  KASCADE-Grande was in operation from 2003 until
2013, after which it was dismantled.

For most investigations, the number of electrons, muons, hadrons is of
interest. These are obtained by sampling the corresponding particle densities
and fitting an empirical function to the particle densities as a function of
distance to the shower axis \cite{kascadelateral}. The integral under these
functions gives the number of electrons, muons, and hadrons in an air shower,
respectively.  They are used in turn to derive cosmic-ray properties, such as
the energy and mass/particle type of the incoming cosmic ray.

\section{Test of hadronic interaction models}
\begin{wrapfigure}{r}{0.40\textwidth}
 \includegraphics[width=0.40\textwidth]{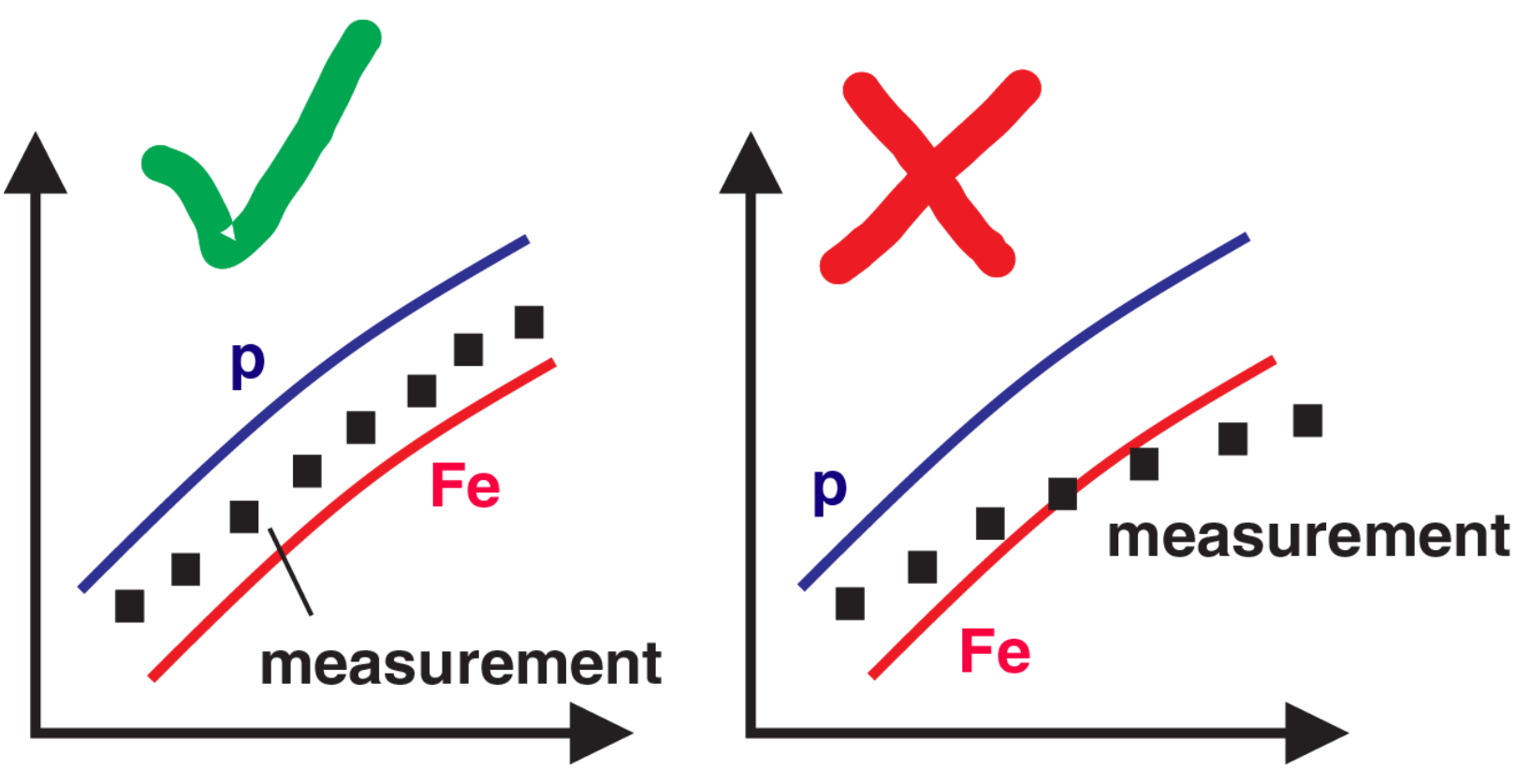}
 \Caption{Principal idea of the feasibility check, see text.}\label{feasibility}
\end{wrapfigure}
The biggest challenge in using cosmic rays to investigate hadronic interactions
is the fact that two properties are (partly) unknown: the precise mass
composition of cosmic rays and the properties of hadronic interactions at
high energies in the forward direction.
To disentangle these problems, the following approach has been adopted
\cite{jrhdurban,Horandel:1998br}:
Cosmic rays comprise all elements known from the periodic table (e.g.\
\cite{cospar06}). However, the abundances of elements heavier than iron
($Z>26$) are significantly lower (by several orders of magnitude) as compared
to the elements with nuclear charge numbers $Z$ up to 26 (hydrogen/protons to
iron). Thus, for a feasibility test, it can be savely assumed that cosmic rays
are mostly comprised of elements from hydrogen/protons to iron.

\begin{figure*}
 \includegraphics[width=0.53\textwidth]{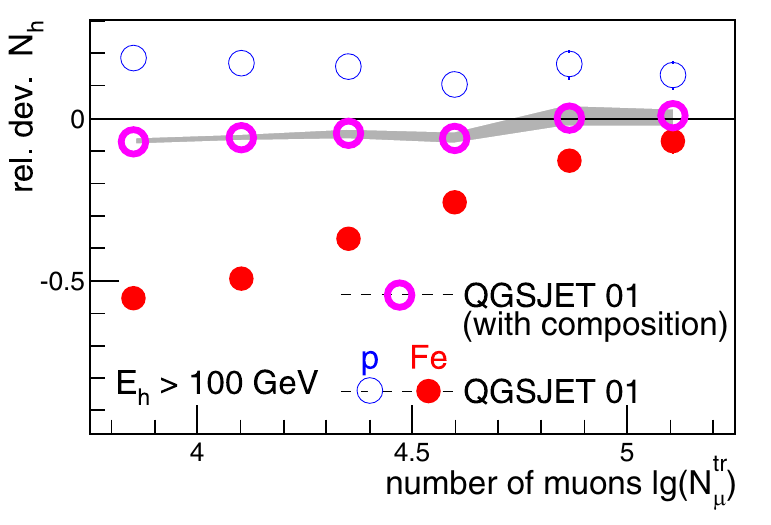}%
 \includegraphics[width=0.47\textwidth]{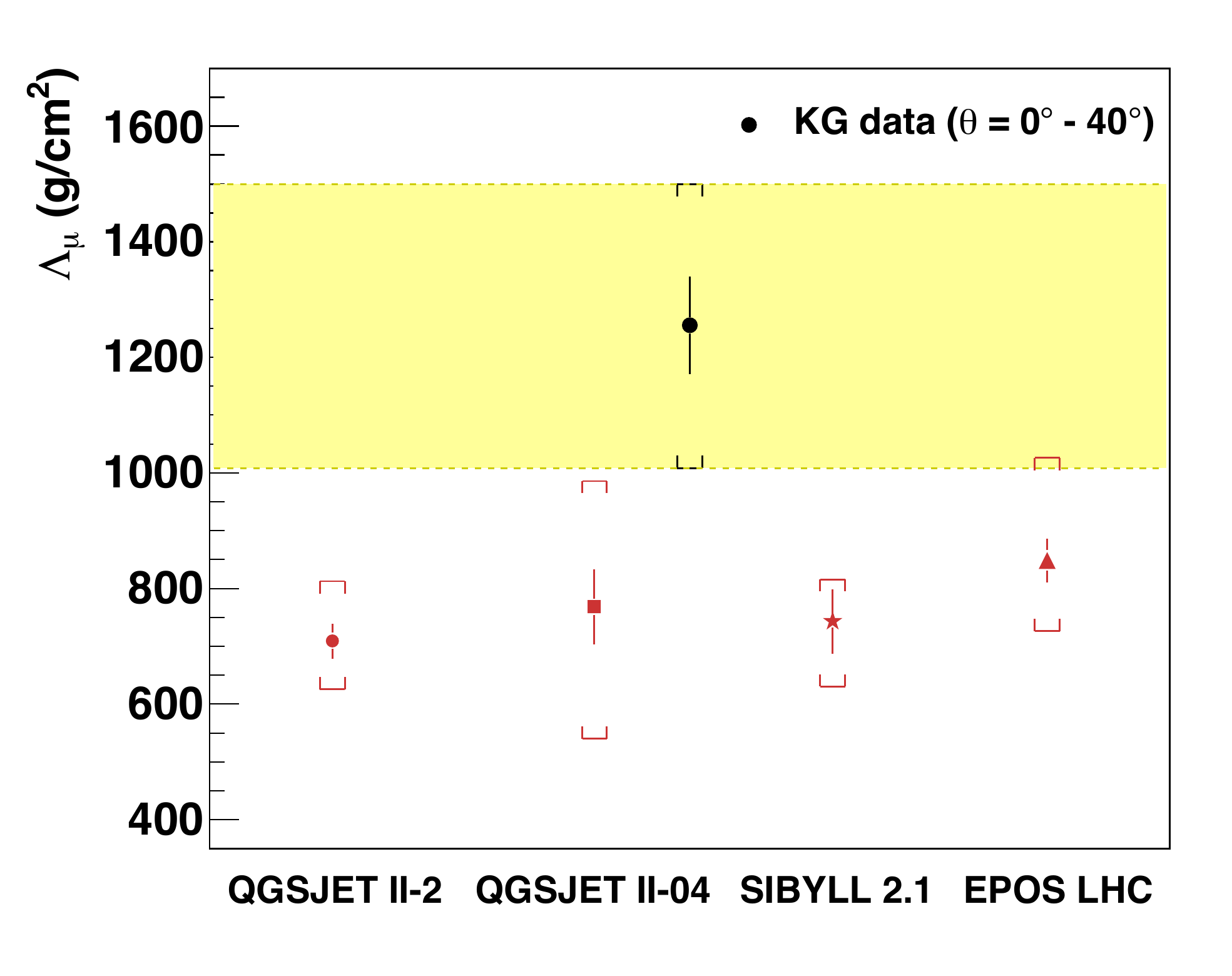}
 \Caption{Left: Number of hadrons at ground level as a function of the number of
muons (integrated over a certain distance range) in air showers. The number of hadrons is given relative to the
expectations of simulations with the model QGSJET~01, assuming a mass
composition as measured with the electromagnetic and muonic detectors of
KASCADE \cite{jenspune}.
 Right: Attenuation length of muons in the atmosphere. Measurements are
  compared to predictions of post-LHC interaction models \cite{attenpap}.}
\label{fig4}
\end{figure*}
Using the extrema protons and iron nuclei, air shower simulations are conducted
with the standard tool CORSIKA \cite{corsika}, using a particular model to
describe the hadronic interactions. The particles reaching ground level are
treated in a detailed detector simulation, based on the standard tool GEANT
\cite{geant}. This results in predictions for the correlation between the
different shower components: electromagnetic, muonic, and hadronic.  These
predictions are compared to measurements of such correlations.  If the
measurements are outside the range protons to iron, a particular hadronic
interaction model is not able to describe the air shower
development consistently, see \fref{feasibility}.

This method has been pioneered by the KASCADE group about two decades ago
\cite{jrhdurban,Horandel:1998br}.
Over time different hadronic interaction models used in air shower simulations
have been studied systematically.
In 1999 the model QGSJET~98 has been favoured over the models VENUS and
SIBYLL~1.6 \cite{wwtestjpg}.
In 2007 the models DPMJET~II.55, QGSJET~01,and SIBYLL~2.1 have been found in
reasonable agreement with the air shower data, while NEXUS~2 exhibited
incompatibilities \cite{jensjpg}.
In 2009 incompatibilities between EPOS~1.6 and air shower data have been found
\cite{epostest} and the model QGSJET~II.2 has turned out to describe the air
shower data best \cite{icrc09-hoerandel}.
It has also been studied how the properties of individual hadronic interactions
(such as the cross section or the multiplicity) affect the overall shower
development \cite{kascadewqpune}.

The mass composition of cosmic rays as a function of energy as obtained from
the measurements of the electromagnetic and muonic shower components (see next
section) has been used to conduct specific air shower simulations to predict
the hadronic component on ground level. Quantitative studies show that the
models QGSJET~01 and SYBILL~2.1 predict the hadronic component with an accuracy
of the order of 10\% \cite{jenspune}, see \fref{fig4} left. 

This work helped to fine-tune the hadronic interaction models used in air
shower simulations.  When first LHC data became available, it was interesting
to realize that the air-shower optimzed hadronic interaction models did quite
well in describing the properties of the first LHC data
\cite{icrc11-pierog,icrc11-sako,icrc11-mitsuka}.

More recenlty, also post-LHC hadronic interaction models have been confronted
with air-shower data.
The influence of current hadronic interaction models on the interpretation of
air shower data has been investigated \cite{icrc13-bertaina}.
The attenuation length of muons in the atmosphere has been measured and
compared to post-LHC hadronic models such as QGSJET~II~04, EPOS-LHC, and
SIBYLL~2.1 \cite{attenpap}. 
While the models predict an attenuation length of the order of 700 to
850~g/cm$^2$, the KASCADE-Grande measurements yield a significantly higher
value between 1200 and 1300 g/cm$^2$, see \fref{fig4} right.  
It is interesting to note that at the Pierre Auger Observatory a similar effect
has been observed at higher energies: more muons are measured as compared to
the predictions of LHC-tuned models \cite{Aab:2016hkv}.

\section{Properties of cosmic rays}
Main objective of KASCADE and KASCADE-Grande is the measurement of the
properties of cosmic rays, in particular to derive energy spectra for elemental
groups in cosmic rays.
\begin{wrapfigure}{r}{0.75\textwidth}
 \includegraphics[width=0.75\textwidth]{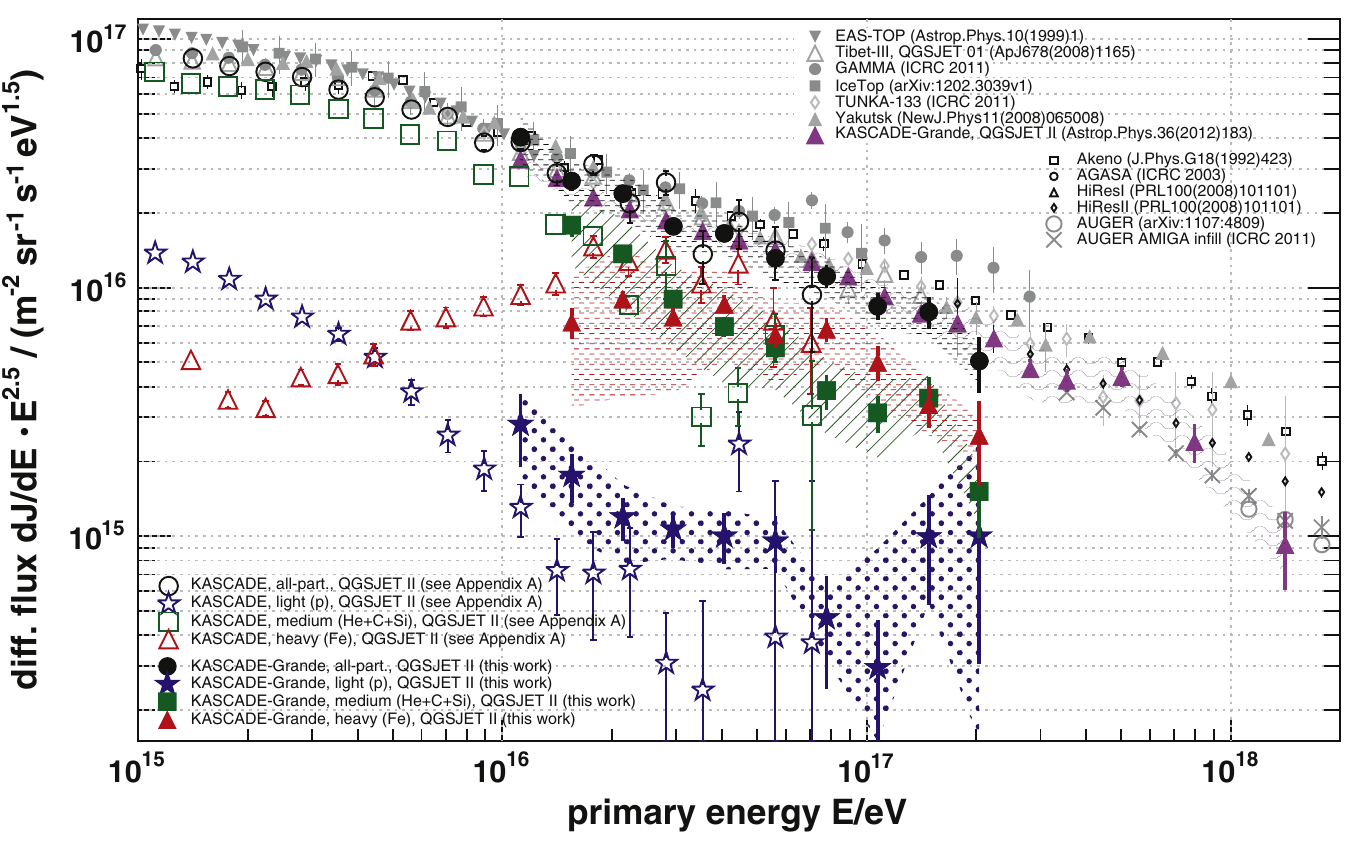}
 \Caption{ The all-particle energy spectrum of cosmic rays. In addition, energy
  spectra for groups of elements are shown (protons/light blue, iron/heavy red),
  for details see \cite{Apel:2013uni}.}\label{espec}
\end{wrapfigure}
The number of electrons and muons in an air shower are obtained through
integration of the lateral density distribution function, as described above.
Applying a simple Heit\-ler-type model, it can be seen that in the electron-muon
number plane the showers are characterized by two axes, an energy axis roughly
alined along the main diagonal and almost perpendicular to it a mass axis
\cite{matthewsheitler,jrherice06}. This illustrates, that from the simultaneous
measurements of the electromagnetic and muonic components the energy spectra
for groups of elements can be unfolded \cite{ulrichapp}. The KASCADE results
indicate that the spectra for elemental groups (protons, helium, CNO group)
follow roughly a power law with a fall off aproximately proportional to the
nuclear charge $Z$. This implies that the knee in the all-particle energy
spectrum of cosmic rays at an energy of about $4-5\cdot10^{15}$~eV is caused by
a fall-off for the light elements. Such a behaviour can be explained by
astrophysical models  (see e.g.\ \cite{origin}): The maximum energy attained
during Fermi acceleration in Supernova Remnants is proportional to $Z$. Also
leakage from the Galaxy during the difussive propagation of cosmic rays through
the Milky Way causes a rigidity dependent fall-off of the spectra.

Following these ideas one would expect a fall-off of the heavy component
(iron-like elements) at an energy $Z_{Fe}=26$ times the energy of the
hydrogen/proton fall-off. Such a fall-off of the heavy cosmic-ray component has
been indeed observed by KASCADE-Grande \cite{Apel:2013uni,Apel:2011mi}. Thus,
confirming the scenario that the individual cosmic-ray elements fall-off at
energies proportional to their nuclear charge $Z$, see \fref{espec}.
It could also be shown that the light component in cosmic rays recovers at
energies above the fall-off energy of the heavy component  \cite{Apel:2013ura}.
Such a behaviour would be expected from a contrbution of another source class
(e.g.\ extragalactic component) at higher energies.

Recently, the IceCube experiment at the South Pole has found a similar
behaviour of a rigidity dependent fall-off of individual elemental groups in
cosmic rays as a function of energy \cite{Rawlins:2016bkc}.
This confirms the findings by KASCADE and KASCADE-Grande and a general picture
emerges (see e.g.\ \cite{Hoerandel:2012wh}): The energy spectra for individual
elements in cosmic rays follow roughly power laws in the GeV and TeV regime. At
higher energies (PeV regime) they exhibit a fall-of proportional to their
nuclear charge $Z$. Thus, the Galactic cosmic-ray component is expected to
reach up to energies above $10^{17}$~eV.

\section{Summary and outlook}
During the last two decades KASCADE and KASCADE-Grande have significantly
increased the knowledge about Galactic cosmic rays at high energies
\cite{Haungs:2015jap}.
Both experiments also improved our understanding of hadronic interactions at
high energies in the kinematic forward direction.
This has been achieved through precise measurements of the individual
components of extensive air showers: the electromagnetic, muonic, and hadronic
components.
The upgraded Pierre Auger Observatory \cite{Aab:2016vlz} with improved
capabilities to measure the electromagnetic and muonic shower components will
allow to continue the quantitative tests of hadronic interaction models with
air shower data, which have been pioneered by KASCADE two decades ago.

\Acknowledgements % if needed
The authors thank the members of the engineering and technical staff of the
KASCADE-Grande collaboration, who contributed to the success of the experiment.
The KASCADE-Grande experiment was supported in Germany by the BMBF and by the
"Helmholtz Alliance for Astroparticle Physics – HAP", funded by the Initiative
and Networking Fund of the Helmholtz Association, by the MIUR and INAF of
Italy, the Polish Ministry of Science and Higher Education, and the Romanian
National Authority for Scientific Research, ANCS-UEFISCDI, project numbers
PN-II-ID-PCE-2011-3-0691 and PN-II-RUPD-2011-3-0145. 
J.C.A.V. acknowledges the partial support of CONACYT (grant CB-2008/106717) and
the German-Mexican bilateral collaboration grants (DAAD-CONACYT 2009–2012,
2015–2016).

{\footnotesize
%\bibliographystyle{elsart-num-jrh}
%\bibliography{cr}

}
 
\end{document}